**Pressure tunable quantum anomalous Hall states in a topological antiferromagnet**

Su Kong Chong[1], Chao Lei[2], Jie Li[3], Yang Cheng[1], David Graf[4], Seng Huat Lee[5,6], Masaki Tanabe[1], Ting-Hsun Yang[1], Zhiqiang Mao[5,6], Allan H. MacDonald[2], Kang L. Wang[1]*

[1]Department of Electrical and Computer Engineering, University of California, Los Angeles, California 90095, United States.

[2]Department of Physics, The University of Texas at Austin, Austin, TX 78712, USA.

[3]School of Materials Science and Engineering, Shanghai University, Shanghai 200444, China.

[4]National High Magnetic Field Laboratory, Florida State University, Tallahassee, FL 32310, USA.

[5]2D Crystal Consortium, Materials Research Institute, The Pennsylvania State University, University Park, PA 16802, USA

[6]Department of Physics, The Pennsylvania State University, University Park, PA 16802, USA

*Corresponding authors: wang@seas.ucla.edu

**Abstract**

**Mechanical modulation of the lattice parameter can modify the electronic structure and manipulate the magnetic coupling of a material without introducing impurities. Inspired by success in pressure-controlled magnetism, we investigate the effect of hydrostatic pressure on quantized Chern states in the antiferromagnetic topological insulator $MnBi_2Te_4$, using transport as a probe. We show that pressure can enhance the robustness of quantum anomalous Hall (QAH) phases that are otherwise delicate in 7SL $MnBi_2Te_4$ and in the spin-flop (SF) state of 8SL $MnBi_2Te_4$. We explain our findings using a coupled Dirac cone model of $MnBi_2Te_4$, which identifies stronger hybridization between van der Waals layers as the driver of topological states. We further demonstrate that moderate pressures readily available in laboratory systems can provide reversible control of**

**magnetic and topological phases. Our results reveal a strong connection between the mechanical engineering of band topology and magnetism.**

**Introduction**

Although intrinsic magnetic topological insulators like $MnBi_2Te_4$ are recognized as promising candidates for high temperature QAH effects,[1-6] progress has been limited by challenges in growing high quality thin films[6] and by the inherent fragility of topological states in these materials. Learning how to prepare more ideal samples is one important research direction. Here we take a different tack by exploring band engineering tools that make the topological states more robust and therefore more tolerant of disorder. Mechanical engineering of crystal lattices has recently received considerable attention as a mechanism for the modulation of electronic and magnetic structures in phases of matter with distinct band topology.[7-14] Pressure can modulate the band structure by (*i*) changing the lattice parameters to control spin-orbit coupling (SOC) strength,[7,8,11,12] and (*ii*) manipulating the magnetic structure to change the exchange coupling between magnetic atoms/spins and Fermi-level electronic states.[15-21] This study is motivated by the sensitivity of interlayer coupling effects in van der Waals materials to pressure and the relationship between interlayer hybridization and topology in $MnBi_2Te_4$ films. Stronger hybridization between layers should lead to more robust topological states.

For layered materials with van der Waals (vdW) gaps, including $MnBi_2Te_4$ in which the gaps separate septuple layer (SL) two-dimensional structural elements, pressure can modify the interlayer separation driving changes in interlayer exchange coupling and hybridization. Previous studies on the A-type antiferromagnet $CrI_3$ [20,21] show that pressure can induce magnetic phase transitions. Compared to $CrI_3$ halide AFMs, $MnBi_2Te_4$ is structurally and magnetically more stable under pressure.[22-25] We nevertheless find that we are able to establish strong control of topological properties with hydrostatic pressure. The hydrostatic pressure is applied by piston cells compatible with electrical transport measurements, while the magnetic and topological properties are monitored by measuring the magnetoresistance ($R_{xx}$) and the Hall resistance ($R_{yx}$) of gated $MnBi_2Te_4$ samples in the Hall bar configuration. The effect of pressure on topological properties in both even and odd layer $MnBi_2Te_4$ can be explained by recognizing their sensitivity to small changes in interlayer hybridization. We establish this connection using first principle electronic structure calculations, Monte Carlo simulations, and approximate coupled Dirac cone models.

**Magnetism under pressure.** The influence of pressure on magnetism in $MnBi_2Te_4$ thin films can be addressed using first-principles electronic structure tools. Fig. 1a illustrates the $MnBi_2Te_4$ atomic structure assumed in our density functional theory (DFT) calculations. By varying magnetic configurations, we constructed an effective Heisenberg spin Hamiltonian for 8SL $MnBi_2Te_4$ (refer to Method) with intralayer ($J_x$) and interlayer ($J_z$) couplings. By systematically changing the pressure applied to the 8SL $MnBi_2Te_4$ thin film, we observe a monotonic increase of $J_z$ with pressure, but an insignificant change in $J_x$, up to 1.6 GPa (Fig. S1), demonstrating that hydrostatic pressure mainly changes the strengths of interlayer exchange interactions. We also investigate the magnetic anisotropy energy (MAE) under pressure for 8SL $MnBi_2Te_4$ in the AFM ground state (refer to Method for details). The results in Fig. 1a show that the MAE, which is due mainly to the contributions from the Bi and Te layers (Fig. S2) and favors perpendicular to plane magnetization,[26] increases slowly with pressure.

In order to understand how external fields couple to the magnetism of $MnBi_2Te_4$ thin films, we perform Monte Carlo numerical simulations using the DFT magnetic-energy landscape[27] to model the pressure dependence of the magnetic states. The magnetic ground state of 8SL MnBi$_2$Te$_4$ under an external field can be solved by minimizing the total magnetic energy. Three transition fields can be identified at what we refer to as the spin-flop ($H_1$), spin-canting ($H_{1'}$), and spin-aligned ($H_2$) fields, where the respective spin structures can be found in Fig. S4b. Their dependence on pressure is summarized in Fig. 1c. The pressure-dependent magnetization and spin orientations in the distinct magnetic phases are shown in Figs. S3 and S4, respectively. The competition between $J_z$, which controls the relative spin orientations in different layers, and the MAE controls the transition fields for field-induced phase transitions.

Fig. 1d shows the Hall resistance ($R_{yx}$) versus magnetic field measured at pressures of 0 GPa and 1.2 GPa for an 8SL MnBi$_2$Te$_4$. The $H_1$, $H_{1'}$, and $H_2$ can be identified from the kinks in the $R_{yx}$ curves [28] as marked by the vertical dashed lines drawn in Fig. 1d (See also Fig. S5). The upward shifts of the magnetic transition fields upon applying pressure are consistent with pressure studies for bulk MnBi$_2$Te$_4$.[22-24] Tracking the transition field at different temperatures, allows the magnetic phase diagram to be mapped. We find reasonable agreement between the experimental magnetic phase transitions and numerical simulations which confirms the enhancement of the interlayer exchange coupling by pressure.

**Spin-flop Chern states in even layer films.** The most intriguing features in our observations are related to field-induced transitions of MnBi$_2$Te$_4$ to Chern insulator states. In the 8SL MnBi$_2$Te$_4$ device we study first, field-induced Hall quantization in the canted antiferromagnetic and ferromagnetic phases shows the necessity of the spin alignment to initiate the Chern insulator state. As shown in Figs. 2a and b, the presence of $R_{yx}$ plateaus at ~$h/e^2$ and full suppression of $R_{xx}$ at magnetic field >5T (<-5T) are indications of the Chern number of C=1 (C=-1) state. The quantization of the Chern insulator in Hall resistivity is comparable to that reported in work in even SL MnBi$_2$Te$_4$.[4,6,29-33]

Comparing the ambient pressure and 0 GPa (device in piston cell) in Figs. 2c and d, the magnetic transition fields do not change, suggesting a negligible pressure effect. The zero Hall plateau ($R_{yx} \approx 0$) due to the AFM coupling at zero (and low) magnetic field is more apparent as the four terminal resistance declines, while the field-induced quantization feature of the Chern insulator states remains. A prominent feature appears in the Chern state quantization upon the application of hydrostatic pressure; at ~1.2 GPa, the system undergoes a steep transition from the AFM insulating state to the Chern insulator state initiated by the SF process. This is revealed in Fig. 2e and f where the $R_{yx}$ reaches quantization of $h/e^2$ together with the sharp suppression of $R_{xx}$ soon as it enters the SF phase. The enhancement in the magnetic transition fields with pressure is consistent with the analyses at higher carrier density (Fig. 1d), suggesting that the pressure-enhanced magnetic exchange coupling remains strong at the exchange gap. This can be better revealed by tuning the gate voltage slightly away from the charge neutrality point (CNP) as shown in Fig. S6, where the SF-induced Chern insulator state remains robust.

To further examine the SF-induced Chern insulator states, we perform field sweeps at different temperatures. The temperature-dependent data for the 0 and 1.2 GPa are compared in Fig. 2c-f where only the positive field sweeps are taken. The temperature response for the anomalous Hall signals in 8SL MnBi$_2$Te$_4$ behaves quite differently in the SF phases as compared to the FM phase. Under 1.2 GPa, the quantization remains robust in the SF phase up to 1.5K and gradually deviates from the quantization value at higher temperatures. In comparison with the 0 GPa case, lowering the temperature does not enhance the anomalous Hall signal any further, suggesting that the ground state in the SF regime is topologically trivial. The topological phase can only emerge at the higher magnetic field when spin alignment enlarges the exchange gap. In the high field region (FM phase), the Chern gap under a pressure of 1.2 GPa is comparable with the 0 GPa Chern gap.

The pressure effect was further checked by removing pressure and measuring the same sample again under ambient pressure. Figs. 2g and h show that similar field-dependent transport behavior is preserved as before pressurizing. The magnetic transition fields and the quantization response return to the initial condition where the fully developed Chern insulator state occurs only in the spin-alignment phase upon releasing pressure. Also, the gate-dependent (Fig. S7) measurements in ambient pressure after pressurization shows no significant difference in the Chern insulator states, indicating that the pressure application can follow a reversible process.

**Emerging QAH phase in odd-layer.** Odd SL $MnBi_2Te_4$ films are expected to be small gap ferromagnets because of their uncompensated layers. We study 7SL $MnBi_2Te_4$ under hydrostatic pressure. We begin by examining the magnetotransport properties of the sample at ambient pressure. As shown in Figs. 3a and b, the high resistance in $R_{xx}$ and $R_{xy}$ color maps and line profiles at zero (and low) magnetic fields suggest an insulating ferromagnetic phase with a non-quantized Hall effect. The hysteretic behavior about zero magnetic field is assigned to the uncompensated layer. The SF field ($H_1$) in odd layer films is higher than in even layer films since the magnetization of the uncompensated layer contributes a finite Zeeman contribution to the total energy under the external magnetic field.[27] At a high magnetic field H>$H_2$, the Chern insulator state emerges due to the alignment of the Mn spins. This is consistent with most reports for odd layer $MnBi_2Te_4$ in literature,[4,31,32] except for the zero field QAH phase observed in 5SL $MnBi_2Te_4$.[6] Applying pressure of 1.2 GPa, we observe strong suppression in the resistive phase at the CNP at zero (and low) magnetic fields as can be seen in the $R_{xx}$ color maps swept around the CNP in Fig. 3c. Also, the field profile of $R_{yx}$ reveals a strong enhancement in Hall hysteresis loop at zero magnetic field (Fig. 3d) with a large coercive field of $H_c \sim 1T$. The anomalous Hall signal reaches a maximum at the CNP as controlled by gate voltage $V_g$ (Fig. S8). The shift in magnetic transition fields $H_1$ and $H_2$ towards higher fields with pressure due to enhancement of the interlayer exchange coupling is consistent with the case of 8SL $MnBi_2Te_4$.

The development of the large anomalous Hall signal indicates an emerging QAH phase when applying hydrostatic pressure to the $MnBi_2Te_4$ antiferromagnet. This is supported by the temperature-dependent $R_{xx}$ and $R_{yx}$ hysteresis loops inserted in Figs. 3c and d, where the anomalous Hall loop continues to develop together with the gradual decrease in $R_{xx}$ with the decrease in temperature. The line profile of $R_{yx}$ reveals a large anomalous Hall loop with a value of $\sim 0.9$ $(-0.7)h/e^2$ observed together with the great suppression in $R_{xx}$ to $\sim 0.2$ $(\sim 0.2)h/e^2$ at 0.5K with sweep-up (sweep-down) field, respectively. Furthermore, the developing QAHI phase can also be indicated by the gate-dependent transport. The $R_{xx}$ and $R_{yx}$ curves as a function of $V_g$ taken at zero magnetic field (Fig. S9) shows a developing $R_{xx}$ minimum at the $R_{yx}$ maximum approaching $h/e^2$ as the Fermi level is tuned to the CNP by the gate voltage. Similar gate-dependent results can be observed for the gate sweep at the higher magnetic field with more robust quantization of the C=1 (C=-1) states developing at negative (positive) magnetic fields due to the complete spin alignment.

The zero (and low) magnetic field in-gap insulating phase reappears upon releasing pressure and the character of the field-induced quantization feature returns to its form prior to pressurization as shown in Figs. 3e and f. This again indicates the reversibility of the transport properties upon subjecting to hydrostatic pressure up to 1.2 GPa, similar to the case for 8SL $MnBi_2Te_4$.

**Pressure-controlled topological phase transitions.** To explain the pressure effect on Chern quantization in even and odd layer MnBi$_2$Te$_4$, we employ a simplified coupled Dirac cone model [34] to calculate electronic structures and topological phases. This model retains only the Dirac cones on the top and bottom surfaces of each septuple layer with the hopping within the septuple layer (across the vdW gap) denoted as $\Delta_S$ ($\Delta_D$). For each Dirac cone, there is an exchange splitting contributed from the magnetic moments of Mn atoms in the same (adjacent) layer denoted as $J_S$ ($J_D$). The details of the calculations are described in the Method section. Our DFT calculations show the dominant change of lattice parameters under hydrostatic pressure happens in physical vdW spacing $d_{vdw}$ with a compression up to ~6% at 1.6 GPa, while the thickness of septuple layer $d_{SL}$ only contracted by <1% at 1.6 GPa (Fig. S10). Following this analysis, we construct the topological phase diagram for the energy gap of 7SL MnBi$_2$Te$_4$ in the dominant parameters, $\Delta_D$, to represent the pressure effect. Since a change of $d_{vdw}$ up to 6% may induce a relative change in $\Delta_D$ that is several times larger than the change in $\Delta_S$, we plotted the topological phase diagram in a reasonable range for $\Delta_D$ and $J_S$ with $J_D$ fixed at $0.8J_S$, as shown in Fig. 4a. This model predicts that odd-layer MnBi$_2$Te$_4$ thin films with a thickness larger 3SL are QAHI (C=1), but defects [35,36] or degradation [37] can significantly reduce the parameter $J_S$, leading to a change in the topology to the trivial insulator (C=0) phase. By assuming about 6% of anti-site defects [35,38], we estimate a reduction of $J_S$ by a factor of 2-3 (Fig. S11), in agreement with the phase diagram in Fig. 4a. Applying pressure can enhance the interlayer Dirac cones coupling $\Delta_D$ as pointed out by the arrow direction in the phase diagram to restore the QAHI phase. The band structure projections constructed by our effective Hamiltonian model from the two pressure tuning points in the phase diagram show clearly the trivial and topological gapped states as shown in Figs. 4a(i) and (ii), respectively. As shown in the phase diagram, the critical point ($\Delta_D^c$) for the pressure-induced topological phase transition line lies between $\Delta_S < \Delta_D < 2\Delta_S$, which can be translated to a change in $d_{vdw}$ of 5-10% (Fig. S12).

Next, we turn to explain the pressure induce topological phase transition in even layer MnBi$_2$Te$_4$. Theoretically, the even layer can either be a trivial AFM insulator or an axion insulator with C=0 at zero field. In the magnetic field-induced spin alignment FM phase, the phase transition leads to the Chern insulator C=1 state with the phase switching point lying in the spin-flip regime. Our experimental results show the sharp transition into the C=1 state happens right at the SF transition upon the application of hydrostatic pressure. The topological phase is clearly distinguished from the same sample without pressure where under the same SF magnetic phase the R$_{yx}$ amplitude does not develop further with the decrease in temperature (refer to Fig. 2). To explore the effect of pressure in 8SL MnBi$_2$Te$_4$ Chern insulator, we construct the LL energy diagrams under the different pressures based on the magnetic moment directions are extracted from our DFT and numerical simulations as illustrated in Figs. 4b(i) and (ii). Comparing the two LL dispersions, we observe a significant enhancement in C=1 Chern insulator gap together with a great suppression in C=0 gap in the SF phase regime under pressure. This SF-induced large C=1 gap at the Fermi level provides strong evidence to support our experimental observation. To further interpret the pressure-induced topological phase transition, we construct a topological phase diagram for the Landau level gap as a function of Chern filling and $\Delta_D$ in the SF phase, as shown in Fig. 4b. The distinct C=0 trivial insulator and the C=1 Chern insulator phases switch at the critical point ($\Delta_D^c$) as a reminiscent of topological phase transition. In contrast, the phase diagram constructed in AFM phase shows a topological phase transition from C=0 AFM insulator to axion insulator states at the critical $\Delta_D^c$, while in the FM phase the critical $\Delta_D^c$ shifts to low energy and thus the C=1 Chern insulator state is easier to realize at high magnetic field even without application of pressure (Fig. S14).

Experimentally, the topological phase transition between the C=0 trivial insulator and the C=1 Chern insulator states can be evaluated from the critical transition regime.[29,38] Owing to their different temperature response, the critical transition field ($B_C$) can be determined from the crossing between the two states. As shown in Figs. 2c and e, the $B_C$ for the 8SL MnBi$_2$Te$_4$ shifts from the spin-canting phase of ~3.2T to SF transition of ~2.4T for 0 GPa and 1.2 GPa, respectively. Meanwhile, the $\rho_{xx}$ converges into a critical resistance value of ~0.8h/e$^2$ at $B_C$ for both the 0 GPa and 1.2 GPa, as shown in Fig. S15.

## Summary

In summary, we studied the magnetoelectrical transport of the intrinsic antiferromagnetic topological insulator MnBi$_2$Te$_4$ under hydrostatic pressure. We showed that the strengthening of the interlayer exchange coupling by pressure can be determined from the shifting in the magnetic transition fields, consistent with our DFT and Monte Carlo-based numerical simulations. We showed that the pressure can also lead to significant enhancement in Chern quantization for thin film MnBi$_2$Te$_4$ as revealed by our well-developed SF Chern insulator state and the emerging QAH insulator phase achieved in the 8SL and 7SL MnBi$_2$Te$_4$, respectively. A careful analysis resolved the distinct topological phases induced by the hydrostatic pressure. Our coupled surface Dirac cone model well-captured similar phenomena by considering the hopping between the adjacent layer as the dominant changing parameter under hydrostatic pressure. More importantly, we found that the pressure control topological phase transition follows a reversible route, opening an opportunity for mechanical tuning topological phase transition in the family of topological antiferromagnets. Our results not only hint to unravel the mystery of the QAH insulator phase in odd layer MnBi$_2$Te$_4$, but the unconventional pressure-controlled magnetic and topological phase transitions in both even and odd layer MnBi$_2$Te$_4$ also pave the path for the realization of the piezo-topological transistors.

**Methods**

**Materials.** MnBi$_2$Te$_4$ bulk crystals were grown by a self-flux growth method.[28] Variable thicknesses of MnBi$_2$Te$_4$ thin flakes were exfoliated from the parent bulk crystals onto the Si/SiO$_2$ substrate which serves as the gate-electrode/dielectric layer. The exfoliation processes were performed in an argon gas-filled glovebox with O$_2$ and H$_2$O levels <1 ppm and <0.1 ppm, respectively, to prevent oxidation in thin flakes. We fabricated the MnBi$_2$Te$_4$ devices into the Hall bar configuration using a standard electron beam lithography process and metal deposition with Cr/Au (20 nm/60 nm) as the contact electrodes using a CHA Solution electron beam evaporator. The MnBi$_2$Te$_4$ flakes were protected by polymethyl methacrylate (PMMA) while transporting for lithography and metal deposition processes.

**Measurements.** Low-temperature and ambient pressure magnetotransport measurements were performed in a Quantum Design Physical Properties Measurement System (PPMS) in helium-4 circulation (2K-300K) and magnetic field up to 9 T. Two synchronized Stanford Research SR830 lock-in amplifiers at a frequency of 5-8 Hertz (Hz) were used to measure the longitudinal and Hall resistances concurrently on the MnBi$_2$Te$_4$ devices. The devices were typically sourced with a small AC excitation current of 10-20 nA. Keithley 2400 source meters were utilized to source DC gate voltages to the bottom Si gate electrodes.

Measurements at variable pressure were conducted in a standard oil-based piston pressure cell. The MnBi$_2$Te$_4$ devices were mounted to the pressure cell using insulating epoxy. The pressure cell was filled with the hydrostatic fluid Daphne 7575 oil. Once assembled, the cell was placed in a hydraulic press where a piston fed through a hole in the threaded top screw of the cell to add pressure. When the appropriate pressure was reached, the top screw was clamped to lock in the pressure. A small ruby chip fixed to the tip of the fiber optic was used to calibrate the pressure at room temperature. The pressure at low temperatures was estimated based on that. The low-temperature and variable pressures magnetotransport measurements were carried out in a helium-3 variable temperature insert at a base temperature of 0.4 Kelvin (K) and magnetic field up to 18 tesla (T) based at the National High Magnetic Field laboratory. As pressure can be varied only at room temperature, the samples went through different thermal cycles during the measurements at different pressures.

**DFT calculations.** The density functional theory calculations were carried out with the Vienna ab-initio simulation package (VASP) at the level of the spin-polarized generalized-gradient approximation (GGA) with the functional developed by Perdew-Burke-Ernzerhof.[39] The interaction between valence electrons and ionic cores is considered within the framework of the projector augmented wave (PAW) method.[40,41] The Hubbard U of 3.0 eV was adopted to describe the electron correlation in the d-shells of Mn cores.[42] The vdW correction (DFT-D3) was included for the description of dispersion forces.[43] The energy cutoff for the plane wave basis expansion is set to 500eV. All atoms are fully relaxed using the conjugated gradient method for the energy minimization until the force on each atom becomes smaller than 0.01 eV/Å, and the criterion for total energy convergence is set at $10^{-5}$ eV.

To analyze the magnetic properties of the 8SL $MnBi_2Te_4$ thin film under hydrostatic pressure, we constructed an effective Heisenberg spin Hamiltonian as

$$H = -J_x \sum_{<i,j>_{intra}} S_i \cdot S_j - J_z \sum_{<a,b>_{inter}} S_a \cdot S_b$$

where $J_x$ and $J_z$ represent the exchange coupling parameters for intralayer and interlayer coupling, respectively. $S_{i,j}$ and $S_{a,b}$ are the spin quantum number for Mn atom (5/2).

The magnetic anisotropy energy (MAE), including the spin-orbit coupling (SOC) driven magnetocrystalline anisotropy (MCA) and the magnetic shape anisotropy energy resulting from dipole-dipole interactions (SAE), with

$$MCA = E_x^{SOC} - E_z^{SOC} = \xi^2 \sum_{u,o,\alpha,\beta} (2\delta_{\alpha\beta} - 1) \left[ \frac{|<u,\alpha|L_z|o,\beta>|^2}{\varepsilon_{u,\alpha} - \varepsilon_{o,\beta}} - \frac{|<u,\alpha|L_x|o,\beta>|^2}{\varepsilon_{u,\alpha} - \varepsilon_{o,\beta}} \right]$$

$$SAE = -\frac{1}{2}\frac{\mu_0}{4\pi} \sum_{m \neq n}^{N} \frac{1}{r_{mn}^3} \left[ \overrightarrow{M_m} \cdot \overrightarrow{M_n} - \frac{3}{r_{mn}^2} (\overrightarrow{M_m} \cdot \overrightarrow{r_{mn}})(\overrightarrow{M_n} \cdot \overrightarrow{r_{nm}}) \right]$$

where ξ is the strength of SOC, $\varepsilon_{u,\alpha}$ and $\varepsilon_{o,\beta}$ are the energy levels of the unoccupied states with spin α and occupied states with spin β, $M_n$ represents the local magnetic moments and $r_{mn}$ are the vectors from the sites *m* to *n*, respectively.

**Numerical simulations.** The numerical simulations for 8-SL MnBi$_2$Te$_4$ were performed by calculating the total magnetic energy as

$$E = \mu_0 M_s \left[ \frac{H_J}{2} \sum_{i=1}^{7} \boldsymbol{m_i} \cdot \boldsymbol{m_{i+1}} - \frac{H_k}{2} \sum_{i=1}^{8} m_{i,z}^2 - H \sum_{i=1}^{8} m_{i,z} \right],$$

where $M_s$ is the saturated magnetization of each sublattice and $\boldsymbol{m_i} = (\sin\theta_i, 0, \cos\theta_i)$ is the unit vector of magnetic moment in *i*th layer with the polar angle $\theta_i$. The total energy contains three terms: interlayer exchange energy with effective exchange field $H_J$, easy axis anisotropy with effective field $H_k$ and Zeeman energy with external field *H*=H(0,0,1). Ground state of 8SL MnBi$_2$Te$_4$ under external field can be solved by minimizing the *E*.[27] For the 8SL MnBi$_2$Te$_4$, when the field increases, the spin-flop transition happens at $H_1$ where the surface layer with magnetic moment antiparallel to the external field flops first. This is also called surface spin flop. As the field increasing, the rest of layers starts to rotate away from easy-axis until the sublattice magnetization staggered aligned with polar angle of each layer $\theta_i = -\theta_{i+1}$. This indicates the start of spin flip ($H_{1'}$) in canted AFM state. Finally, when the Zeeman energy overcomes the antiferromagnetic exchange energy, all $\theta_i = 0$, MnBi$_2$Te$_4$ gets into FM state ($H_2$). The $H_1$ and $H_2$ are determined by the magnitude of $H_J$ and $H_k$ as

$$H_1 \sim \sqrt{H_J H_K - (H_K)^2}$$

$$H_2 = 2\cos^2\left(\frac{\pi}{16}\right) H_J - H_K \sim 2H_J - H_K$$

Our best fit to the experimental data gives $H_J = 4.4\,T$ and $H_K = 1.1\,T$ at 0 GPa, which is consistent with previous report.[27]

**Effective Hamiltonian models.** The effective model retains only the Dirac cone surface states on the top and bottom surfaces of each septuple layer and hoppings between these Dirac cones. For each Dirac cone, there is an exchange splitting contributed from the magnetic moments of Mn atoms. With these ingredients the Hamiltonian reads:

$$H = \sum_{k_\perp, ij} \left[ ((-)^i \hbar v_D (\hat{z} \times \sigma) \cdot k_\perp + m_i \sigma_z) \delta_{ij} + \Delta_{ij}(1 - \delta_{ij}) \right] c^\dagger_{k_\perp i} c_{k_\perp j},$$

where $i$ and $j$ label the Dirac cone, $\hbar$ is the reduced Planck constant, and $v_D$ is the velocity of the Dirac cones. $m_i = \sum_\alpha M_\alpha J_\alpha$ is the exchange splitting of the $i$th Dirac cone with $M_\alpha = \pm 1$ the direction of magnetic moments forms the Mn atoms of $\alpha$ layer. The hopping between the $i$th and $j$th surface is denoted by $\Delta_{ij}$. Here we only retain the most dominant parameters, i.e., the hopping between Dirac cones within the same layer $\Delta_S$ and between adjacent layers $\Delta_D$, the exchange splittings only consider the contributions from the magnetic moments of Mn atoms in the same layer $J_S$ and from adjacent layers $J_D$. These parameters in the model can be estimated by comparing the band energies from the couple Dirac cone model and DFT calculations.

In the present of perpendicular external magnetic field $\boldsymbol{B} = B\hat{z}$ and choose the Landau gauge $\boldsymbol{A} = (0, -Bx, 0)$, the quantized Dirac Hamiltonian becomes:

$$H_D = \frac{1}{2}\hbar\omega_c(\sigma^+ a^\dagger + \sigma^+ a)$$

where $\sigma^\pm = \sigma_x \pm \sigma_y$, $a = 1/\sqrt{2}(\tilde{x} + \partial\tilde{x})$, $a^\dagger = 1/\sqrt{2}(\tilde{x} - \partial\tilde{x})$ and $\omega_c = \hbar v_D/\sqrt{2}l_B$. Here $l_B = \sqrt{hc/eB}$ is the magnetic length, $\tilde{x} = l_B k - x/l_B$ and $\partial\tilde{x} = l_B \partial_x$. The general wavefunction in Landau level representation is expressed as

$$|n, i\sigma\rangle = \sum_j \left(C_{nj\uparrow}|n, j\uparrow\rangle + C_{nj\downarrow}|n-1, j\downarrow\rangle\right)$$

with i,j represents the surface indexes, $\sigma = \uparrow/\downarrow$ labels the spin and n = 0, 1,… is the Landau level index. Instead of a gap closing across the topological phase transition in the absence of magnetic field, the Landau levels appear when a magnetic field is applied, the changes of Chern numbers happen each time the Landau levels are crossed.

**Data Availability**

The data that support the findings of this study are available from the corresponding authors upon reasonable request.

**Acknowledgments.**

This work was supported by the National Science Foundation the Quantum Leap Big Idea under Grant No. 1936383 and the U.S. Army Research Office MURI program under Grants No. W911NF-20-2-0166 and No. W911NF-16-1-0472. Support for crystal growth and characterization was provided by the National Science Foundation through the Penn State 2D Crystal Consortium-Materials Innovation Platform (2DCC-MIP) under NSF cooperative agreement DMR-2039351. A portion of this work was performed at the National High Magnetic Field Laboratory, which is supported by National Science Foundation Cooperative Agreement No. DMR-1644779 and the State of Florida.

**Author Contributions**

S.K.C. and K.L.W. planned the experimental project. S.H.L. and Z.M. prepared the bulk crystals. M.T and T.H.Y. helped with the sample preparation. S.K.C. fabricated the devices and conducted the transport measurements. D.G. helped with the piston cell setup and measurements conducted at the National High Magnetic Field laboratory. C.L. and A.H.M. performed calculations on the effective Hamiltonian models. J.L. performed the DFT studies. Y.C. performed the Monte Carlo simulations. S.K.C., L.C., J.L., Y.C., A.H.M., and K.L.W. wrote the manuscript. All authors discussed the results and commented on the manuscript.

**Figures**

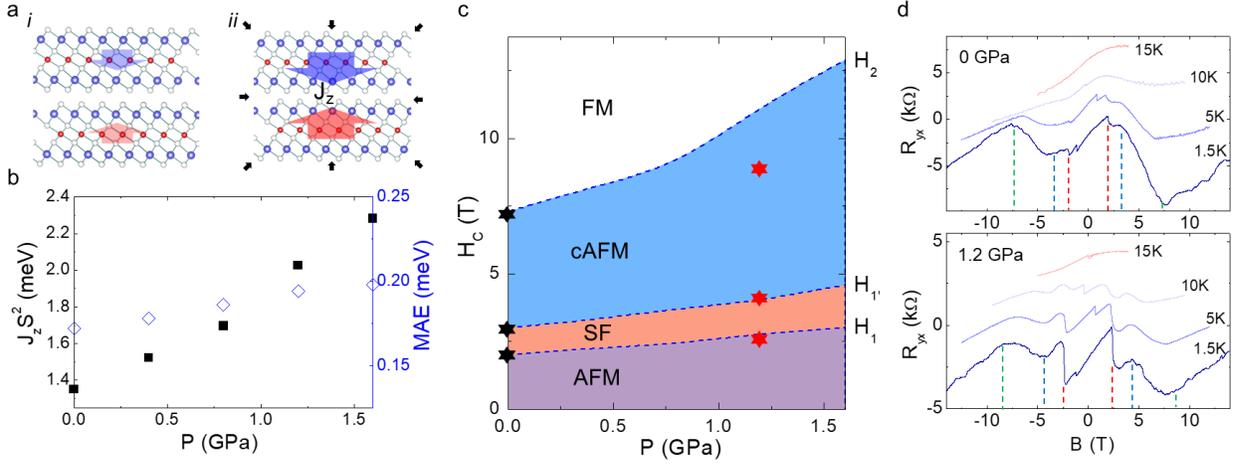

**Figure 1. Magnetic properties of MnBi$_2$Te$_4$ modulated by pressure.** (a) Schematic of the crystal structure of MnBi$_2$Te$_4$ under (*i*) ambient and (*ii*) applied pressures. (b) DFT calculated exchange coupling along the c-axis ($J_z$) and the magnetic anisotropy energy (MAE) as a function of pressure. (c) Calculated magnetic transition fields, $H_1$, $H_{1'}$ and $H_2$, obtained from numerical simulations using the parameters obtained from DFT. The AFM, SF, cAFM, and FM denote the antiferromagnetic, spin-flop, canted antiferromagnetic, and ferromagnetic phases, respectively. The black and red dots in (c) represent the experimental data of the magnetic transition fields extracted from $R_{yx}$ versus magnetic field curves under 0 and 1.2 GPa, respectively. (d) $R_{yx}$ as a function of magnetic field for the 8SL MnBi$_2$Te$_4$ under (*i*) 0 and (*ii*) 1.2 GPa measured at different temperatures at gate voltage controlled into hole carrier region. The $R_{yx}$ curves obtained at different temperatures were manually shifted upward for clarity. The red, blue and green vertical dashed lines in (d) trace the $H_1$, $H_{1'}$ and $H_2$, respectively, as determined from the kinks in the $R_{yx}$ curves.

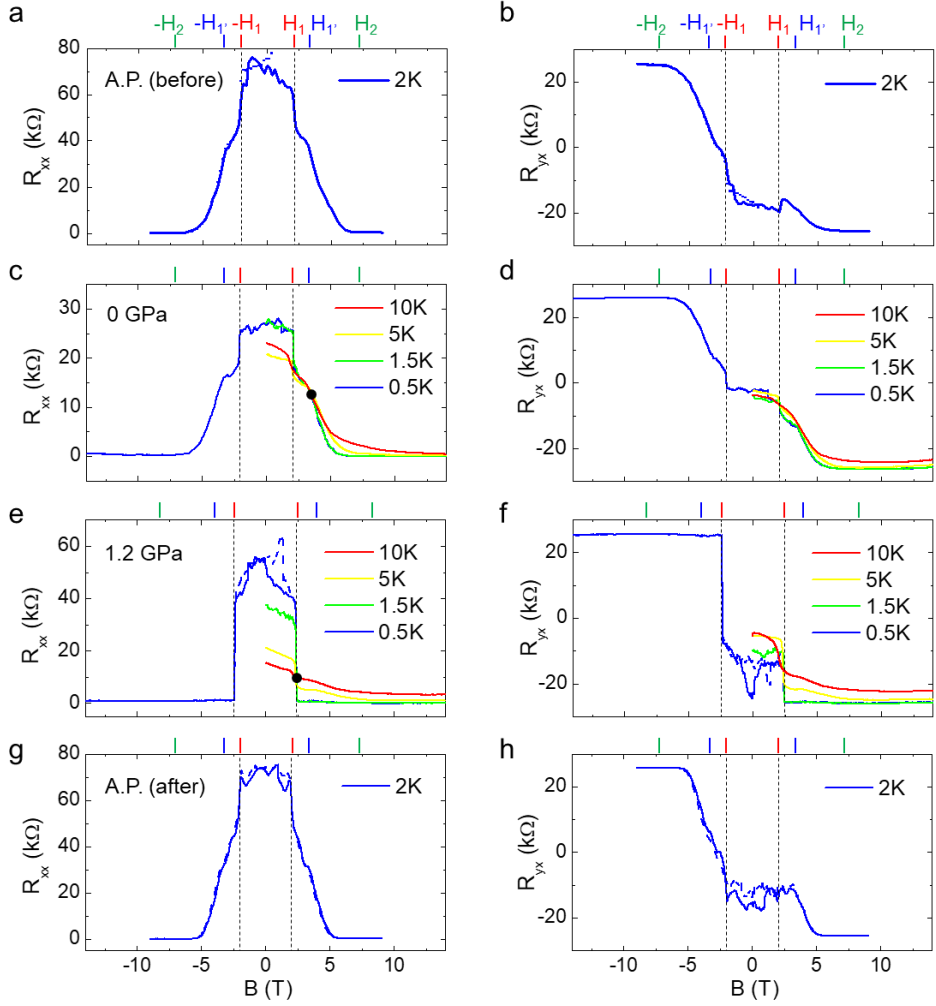

**Figure 2. Even layer MnBi₂Te₄ under pressure.** (a) $R_{xx}$ and (b) $R_{yx}$ as a function of magnetic field taken at CNP ($V_g=V_g^0$) for an 8SL MnBi$_2$Te$_4$ under different conditions: A.P. is ambient pressure, 0 GPa is under negligible pressure in a piston, 1.2 GPa is under pressurized condition. The A.P. or ambient pressure conditions were measured in an in-house PPMS system (base temperature 2K), while the 0 and 1.2 GPa in a helium-3 insert (base temperature ~0.5K). The line indicators trace the magnetic transition fields for spin-flop ($H_1$), cAFM ($H_{1'}$), and FM ($H_2$) transitions, respectively, under different pressure conditions. The $R_{xx}$ and $R_{yx}$ field sweeps under different temperatures for the 0 GPa and 1.2 GPa are included for comparison. Solid and dashed lines for the same color represent the field sweep in a forward and backward direction, respectively. The color lines represent the field sweep at different temperatures. The black dot in (c) and (e) represents the critical point for the transition between C=0 and C=-1 states.

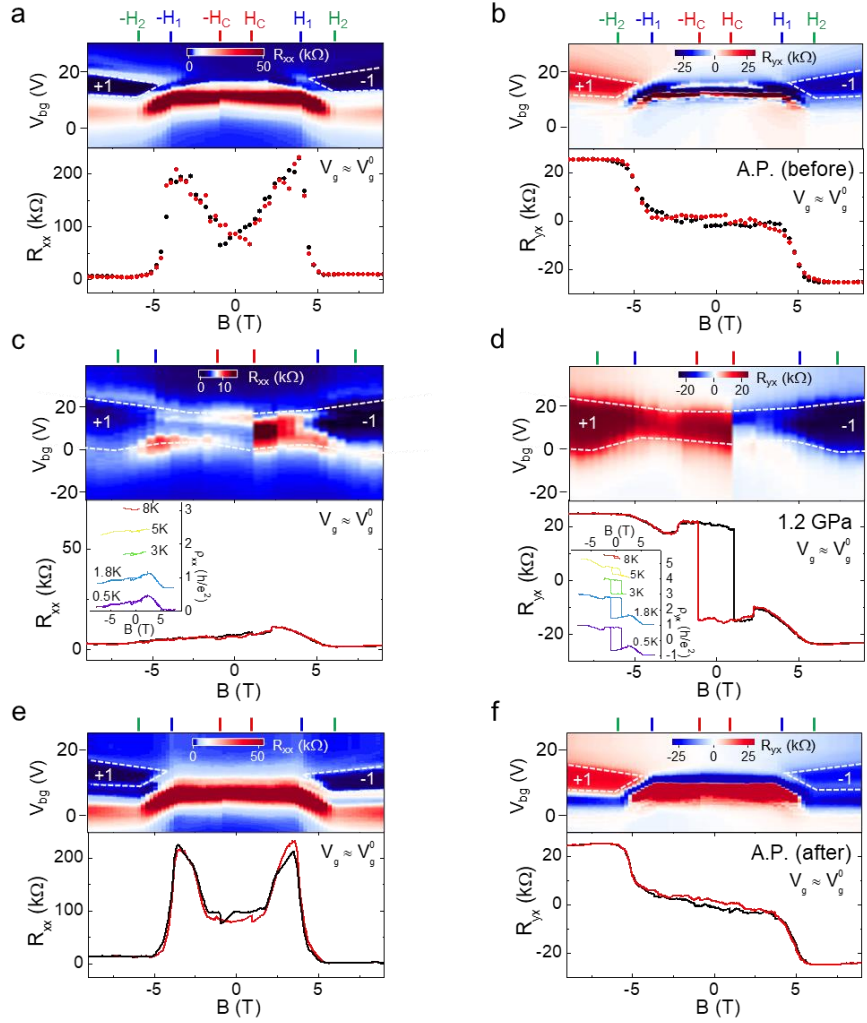

**Figure 3. Odd layer MnBi$_2$Te$_4$ under pressure.** Color maps (top) of the R$_{xx}$ and R$_{yx}$ as a function of magnetic field and gate voltage for a 7SL MnBi$_2$Te$_4$ measured under different pressure conditions, as (a,b) A.P. (before pressurization), (c,d) 1.2 GPa, and (e,f) A.P. (after pressurization). The A.P. or ambient pressure conditions were measured in an in-house PPMS system (base temperature 2K), while the 1.2 GPa in a helium-3 insert (base temperature ~0.5K). Line profiles (bottom) of the respective R$_{xx}$ and R$_{yx}$ versus magnetic field taken at the charge neutrality under different pressure conditions. The R$_{yx}$ curves in (b) and (f) were antisymmetrized to remove the resistive background. Line indicators at the top trace the coercive and transition fields under different pressure conditions. White dashed lines in the R$_{xx}$ and R$_{yx}$ color maps trace the region for the C=1 and -1 states. Insets in (c) and (d) are the magnetic field dependent R$_{xx}$ and R$_{yx}$, respectively, measured at different temperatures under a pressure of 1.2 GPa.
22

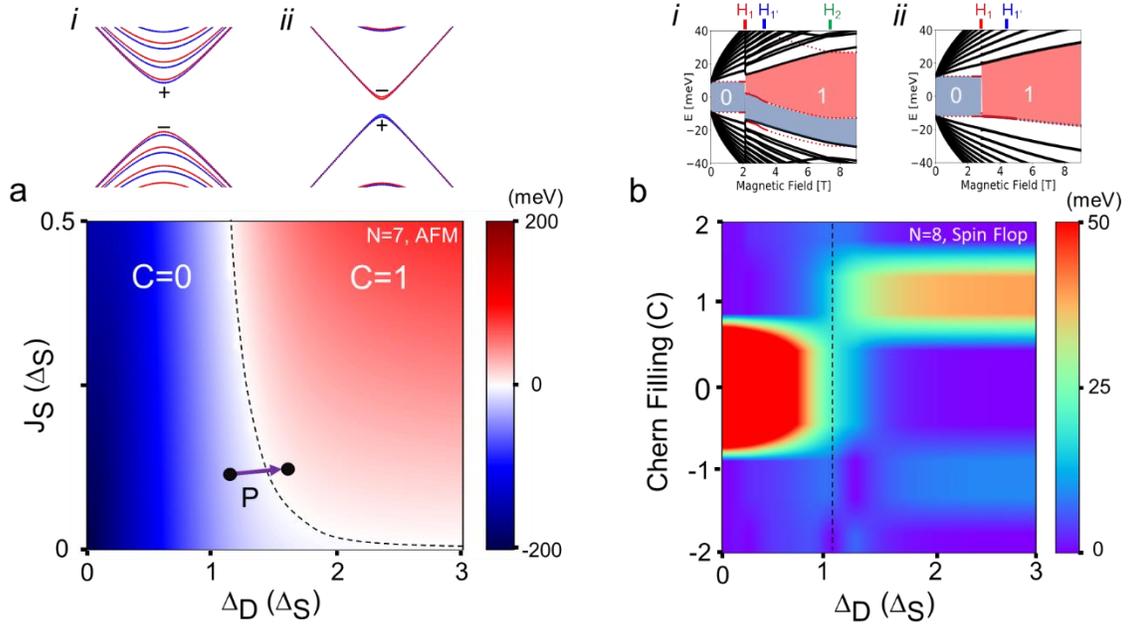

**Figure 4. Pressure-induced topological phase transitions.** (a) Topological phase diagram as a function of $J_S$ and $\Delta_D$ for the trivial insulator (C=0) and QAHI (C=1) phases for 7SL $MnBi_2Te_4$ constructed at zero magnetic field (B=0). The color scale represents the size and parity of the exchange gap. The dashed line in (a) indicates the criticality at the topological phase transition. The arrow in (a) points to the direction of tuning with hydrostatic pressure. Band structure projections for (i) P=0 and (ii) P>$P_c$ represent the distinct trivial and topological phases, respectively. The trivial and topological insulator states are differentiated by the parity (+) and (-). (b) Topological phase diagram as a function of Chern filling and $\Delta_D$ for 8SL $MnBi_2Te_4$ constructed at the SF transition field (H=$H_1$). Dashed line in (b) represents the critical $\Delta_D^c$ for the topological phase transition in the SF regime. The Landau level energy spectra of the 8SL $MnBi_2Te_4$ constructed as a function of Fermi energy and magnetic field under the different pressure conditions of (i) P= 0, and (ii) P= 1.2 GPa. The blue and red shades denote the gaps with Chern number of C=0 and 1, respectively.